\documentclass[fleqn,usenatbib]{mnras}

\usepackage{newtxtext,newtxmath}
\usepackage[T1]{fontenc}
\usepackage{ae,aecompl}
\usepackage{placeins}

\usepackage{color}

\usepackage{graphicx}	% Including figure files
\usepackage{amsmath}	% Advanced maths commands
\usepackage{amssymb}	% Extra maths symbols

\def\msunh{\,h^{-1}\rm{\,M_{\odot}}}
\def\msun{\rm{\,M_{\odot}}}

\def\rtwo{R_{\rm 200}}
\def\kms{\rm{\,km\,s^{-1}}}
\def\repose{$\rm NoAcc\Delta t$}

\def\mw{{\rm m-weight}}
\def\rw{{\rm r-weight}}
\def\mrw{{\rm mr-weight}}
\def\now{{\rm no-weight}}

\definecolor{orange}{rgb}{1,0.5,0}

\definecolor{amethyst}{rgb}{0.6, 0.4, 0.8}
\definecolor{ao(english)}{rgb}{0.0, 0.5, 0.0}

\title[BCG-Cluster Alignment]{Evolution and Role of Mergers in the BCG-Cluster Alignment. A View from Cosmological Hydro-Simulations}
\author[C. Ragone-Figueroa et al.]{
C. Ragone-Figueroa$^{1,2}$,\thanks{E-mail: cinthia.ragone@unc.edu.ar}
G. L. Granato$^{2,1,3}$,
S. Borgani$^{2,3,5,6}$,
R. De Propris$^{4}$,\\~\\
{\rm {\LARGE  D. Garc\'ia Lambas$^{1,2}$,
G. Murante$^2$,
E. Rasia$^{2,3}$ and
M. West$^7$
}}
\\
\\
$^{1}$ Instituto de Astronom\'ia Te\'orica y Experimental (IATE), Consejo Nacional de Investigaciones Cient\'ificas y T\'ecnicas de la\\ Rep\'ublica Argentina (CONICET), Universidad Nacional de C\'ordoba, Laprida 854, X5000BGR, C\'ordoba, Argentina\\
$^{2}$ INAF, Osservatorio Astronomico di Trieste, via Tiepolo 11, I-34131, Trieste, Italy \\
$^{3}$ IFPU - Institute for Fundamental Physics of the Universe, Via Beirut 2, 34014 Trieste, Italy\\
$^{4}$ FINCA, University of Turku, Vaisalantie 20, Piikkio FI-21500, Finland\\
$^5$ Dipartimento di Fisica dell' Universit\`a di Trieste, Sezione di Astronomia, via Tiepolo 11, I-34131 Trieste, Italy
\\
$^6$ INFN - National Institute for Nuclear Physics, Via Valerio 2, I-34127 Trieste, Italy\\
$^7$ Lowell Observatory, 1400 W Mars Hill Rd, Flagstaff, AZ 86001, USA \\
}

\date{Accepted XXX. Received YYY; in original form ZZZ}

\pubyear{2020}

\begin{document}
\label{firstpage}
\pagerange{\pageref{firstpage}--\pageref{lastpage}}
\maketitle
% Abstract of the paper
\begin{abstract}
Contradictory results have been reported on the time evolution of the alignment between clusters and their Brightest Cluster Galaxy (BCG).
We study this topic by analyzing cosmological hydro-simulations of 24 massive clusters with $M_{200}|_{z=0} \gtrsim 10^{15}\, \msun$, plus 5 less massive with $1 \times 10^{14} \lesssim M_{200}|_{z=0} \lesssim 7 \times 10^{14}\, \msun$, which have already proven to produce realistic BCG masses.
We compute the BCG alignment with both the distribution of cluster galaxies and the dark matter (DM) halo. At redshift $z=0$, the major axes of the simulated BCGs and their host cluster galaxy distributions are aligned on average within 20$^\circ$. The BCG alignment with the DM halo is even tighter. The alignment persists up to $z\lesssim2$ with no evident evolution.  This result continues, although with a weaker signal, when considering the projected alignment.
The cluster alignment with the surrounding distribution of matter ($3R_{200}$) is already in place at $z\sim4$ with a typical angle of $35^\circ$, before the BCG-Cluster alignment develops. The BCG turns out to be also aligned with the same matter distribution, albeit always to a lesser extent. These results taken together might imply that the BCG-Cluster alignment occurs in an outside-in fashion.
Depending on their frequency and geometry, mergers can promote, destroy or weaken the alignments.  Clusters that do not experience recent major mergers are typically more relaxed and aligned with their BCG. In turn, accretions closer to the cluster elongation axis tend to improve the alignment as opposed to accretions closer to the cluster minor axis.
\end{abstract}

% Select between one and six entries from the list of approved keywords.
% Don't make up new ones.

%%%%%%%%%%%%%%%%%%%%%%%%%%%%%%%%%%%%%%%%%%%%%%%%%%

%%%%%%%%%%%%%%%%% BODY OF PAPER %%%%%%%%%%%%%%%%%%
\section{Introduction}
It has long since been established that in the local Universe brightest cluster galaxies (BCGs) tend to be elongated in the same direction as their host clusters, as originally noted by \cite{sastry1968}.
Later work has demonstrated that this general alignment is detectable independently of the particular tracer of the cluster shape, such as the distribution of member galaxies \citep[e.g.][]{niederste-ostholt2010}, the X-Ray emitting hot gas or the Sunyaev-Zeldovich effect \citep{hashimoto2008, donahue2016}, or the total mass maps derived from strong and weak lensing  \citep{donahue2016}. Recently, it has also been pointed out that the large-scale environment may also play an important role \citep{wang2018}.

Besides the BCG-Cluster alignment, several other forms of preferred orientation of cosmic structures have been investigated \cite[for a review see][]{joachimi2015}, such as the BCG alignment with large-scale structures \citep[e.g.][]{argyres1986,lambas1988}, or the still controversial tendency for major axes of cluster satellite galaxies to point toward the cluster center \citep[see][and references therein]{huang2016}, or the correlation between cluster shape and large-scale structures \citep[e.g.][]{ragone2007,paz2011}.
However, in this work we will concentrate only on the first, well established, phenomenon and its dependence on redshift.

The redshift dependence of the BCG-Cluster alignment is comparatively scanty known, and different authors reported  contradictory results already at moderate redshift $z \lesssim 0.4$. Indeed, while both
\cite{niederste-ostholt2010} and \cite{hao2011} claim that the alignment signal becomes weaker at higher redshift, the same trend was not confirmed later by \cite{huang2016}. At yet higher redshift an impressive result was reported by \citet{west2017}, who found a clear BCG-cluster alignment signal for ten clusters at $z > 1.3$ observed with the Hubble Space Telescope.

In principle, some theoretical insight on the origin of the BCG-Cluster alignment can be attained
by means of gravity-only cosmological simulations, comparing the direction of the major axes of dark
matter (DM) haloes at different scales \citep[see][and references therein]{kang2007, suto2016}. However, reliable investigations require hydro-dynamical simulations, including the
sub-resolution description of physical processes (e.g. star formation and feedback effects) which are necessary to produce BCGs as close a possible to the observed population. Recently, in \citet{ragone2018}
we have shown that our zoom-in simulations of 24 massive galaxy cluster predict a mass growth and a final mass of BCGs in reasonable agreement with available observational results. Therefore we devote this paper to investigate the evolution of the BCG-cluster alignment, as predicted by the same simulations.

Most previous analyses of the BCG-Cluster alignment based on cosmological hydrodynamical simulations \cite[e.g.][]{dong2014,Velliscig2015,tenneti2015,zhang2019,okabe2019} have been focused on smaller mass clusters ($\lesssim \mbox{a few} \, 10^{14} \mbox{M}_\odot$), selected in significantly smaller volumes than that of our parent simulation. Moreover, in this work, we are specifically interested in quantifying the role of major mergers (mass ratio $>0.25$) on the alignment. On the observational side, recently \cite{wittman2019} claimed that the alignment distribution of clusters undergoing major mergers, around 1 Gyr after the first pericenter passage, is consistent with that of the general population. Note that this conclusion is based on the assumption that the direction connecting the two BCGs is a proxy for that of the filament along which the clusters are merging, as well as the major axis of the eventual merged cluster. Taken at face value, their finding suggests that any plausible worsening of the alignment caused by the merger should fade quickly.

The organization of the present paper is as follows. In Section \ref{sec:numsim} we summarize the characteristics of our simulations. The analysis method is described in the subsequent Section \ref{sec:method}, and the results are presented in Section \ref{sec:result}. The final Section \ref{sec:conclusion} summarizes our main conclusions.

%For higher z alignments:
%\citet{hashimoto2008} find a strong alignment signal using a sample of ....
%no entiendo bien como miden el alignment xq describen una muestra de 120 cumulos X 0.05<z<1.26 y una de $30$ BCG with redshifts between $0.08$ and $0.9$. Cluster and BCG shape were measured using X-ray and optical data respectively. En este paper no dicen nada acerca de que haya una evolucion del alignment con el z, simplemente tienen una muestra que va al menos hasta z0.9

\section{Numerical Simulations}
\label{sec:numsim}
The numerical simulations analyzed in this paper are presented in \cite{ragone2018}. These simulations are similar to those presented in \cite{ragone2013}, but include an updated version of the AGN feedback scheme. Therefore, here we only describe their most relevant features for the present study.  For further numerical or technical details on this set of simulations, we refer the reader to the above papers, and references therein.

Our set consists of 29 zoomed-in Lagrangian regions with a custom version of the {\footnotesize GADGET-3} code~\cite[][]{springel2005}. These regions have been selected from a parent gravity-only simulation of a 1 $h^{-1}$Gpc box, and are centered around the 24 most massive dark matter (DM) haloes. They all have masses\footnote{$M_{200}$ ($M_{500}$) is the mass enclosed by a sphere whose mean density is 200 (500) times the critical density at the considered redshift. The radius of this sphere is dubbed $R_{200}$ ($R_{500}$)} $M_{200} \gtrsim 1.1 \times 10^{15}\, \msun$. In addition we select randomly 5 less massive haloes with masses $1.4 \times 10^{14} \lesssim M_{200} \lesssim 6.8 \times 10^{14}\, \msun$. Each region was re-simulated at higher resolution including hydrodynamics and sub-resolution baryonic physics.
The adopted cosmological parameters are: $\Omega_{\rm{m}} = 0.24$, $\Omega_{\rm{b}} = 0.04$, $n_{\rm{s}}=0.96$, $\sigma_8 =0.8$ and $H_0=72\,\kms$\,Mpc$^{-1}$. The mass resolution for the DM and gas particles is $m_{\rm{DM}} = 8.47\times10^8 \, \msunh$ and $m_{\rm{gas}} =1.53\times10^8\, \msunh$,
respectively. For the gravitational force, a Plummer-equivalent softening length of $\epsilon = 5.6\, h^{-1}$\,kpc is used for DM and gas particles, whereas $\epsilon = 3\, h^{-1}$\,kpc for black hole and star particles. The DM softening length is kept fixed in comoving coordinates for $z>2$ and in physical coordinates at lower redshift.

%we used the SPH formulation by \cite{beck2016}, that includes artificial thermal diffusion and a higher-order interpolation kernel, which improves the standard SPH performance in its capability of treating discontinuities and following the development of gas-dynamical instabilities.

Our set of simulations includes a treatment of several sub-resolution baryonic processes usually included in galaxy formation simulations. For details on the adopted implementation of cooling, star formation, and associated feedback, we refer the reader to \cite{ragone2013}. %In brief, the model of SF is an updated version of the implementation by~\cite{springel2003}, in which gas particles with a density above $0.1\,$cm$^{-3}$ and a temperature below $2.5\times 10^5$\,K are classified as multiphase. Multiphase particles comprise a cold and a hot-phase, in pressure equilibrium. The cold phase is the star formation reservoir.
Metallicity dependent cooling is implemented following the approach by \cite{wiersma2009}. The production of metals is followed according to the model of stellar evolution originally implemented by \cite{tornatore2007}.

A full account of the AGN feedback model can be found in Appendix A of \cite{ragone2013}, with few modifications discussed in Section 2 of \cite{ragone2018} and required to improve the spatial association of the particles representing SMBH  with the stellar system in which they were first seeded. This is fundamental to obtain the best possible effect of AGN feedback in limiting the stellar mass growth. The same set of simulations has been used in \cite{bassini2019} to study SMBH-cluster scaling relations.

Throughout the paper, comoving distances will be denoted by the c letter put before the corresponding unit, that is cMpc for comoving Mpc. Otherwise, we are referring to physical distances.

\begin{figure*}
    \includegraphics[width=5.5cm]{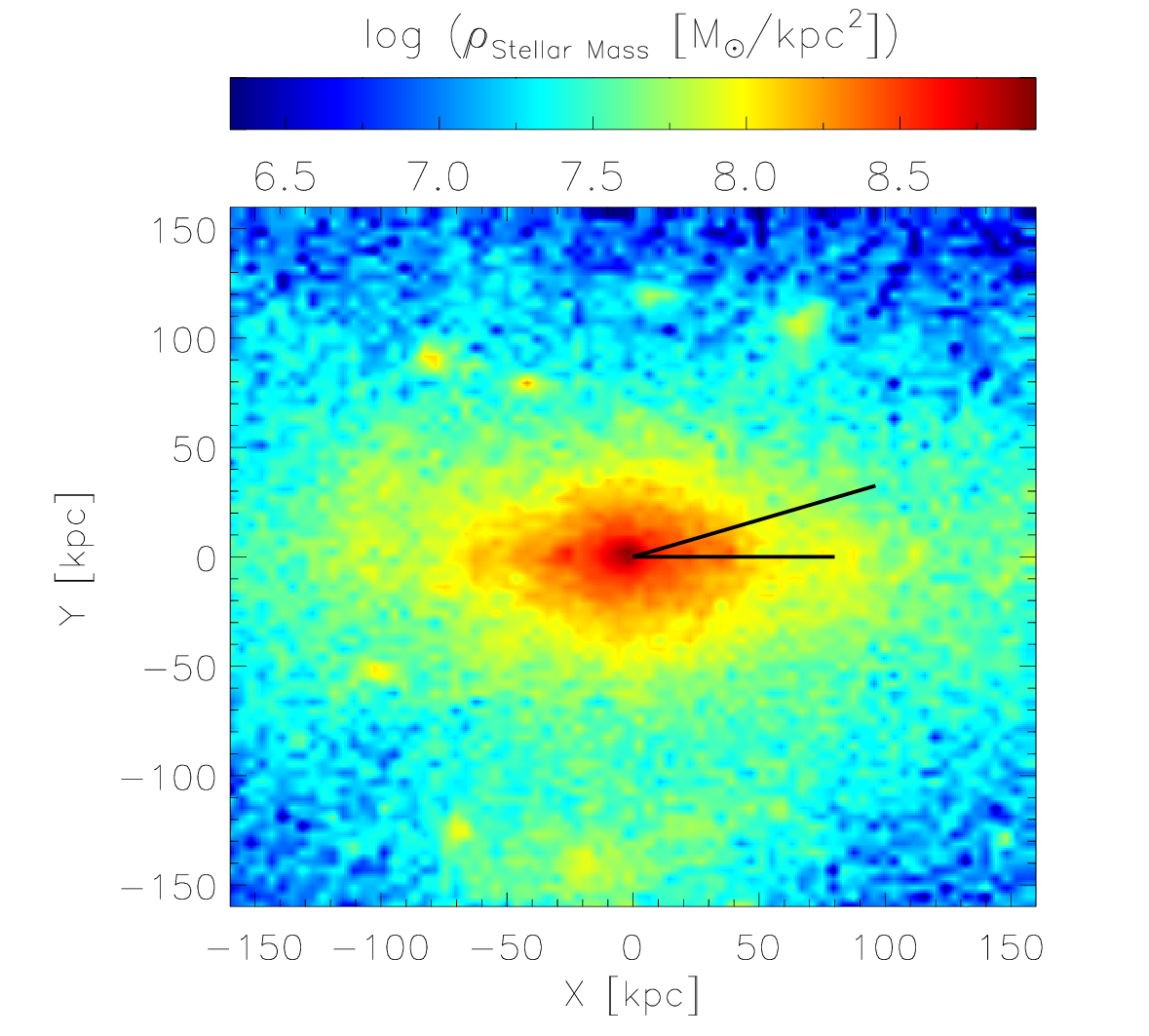}
    \includegraphics[width=5.5cm]{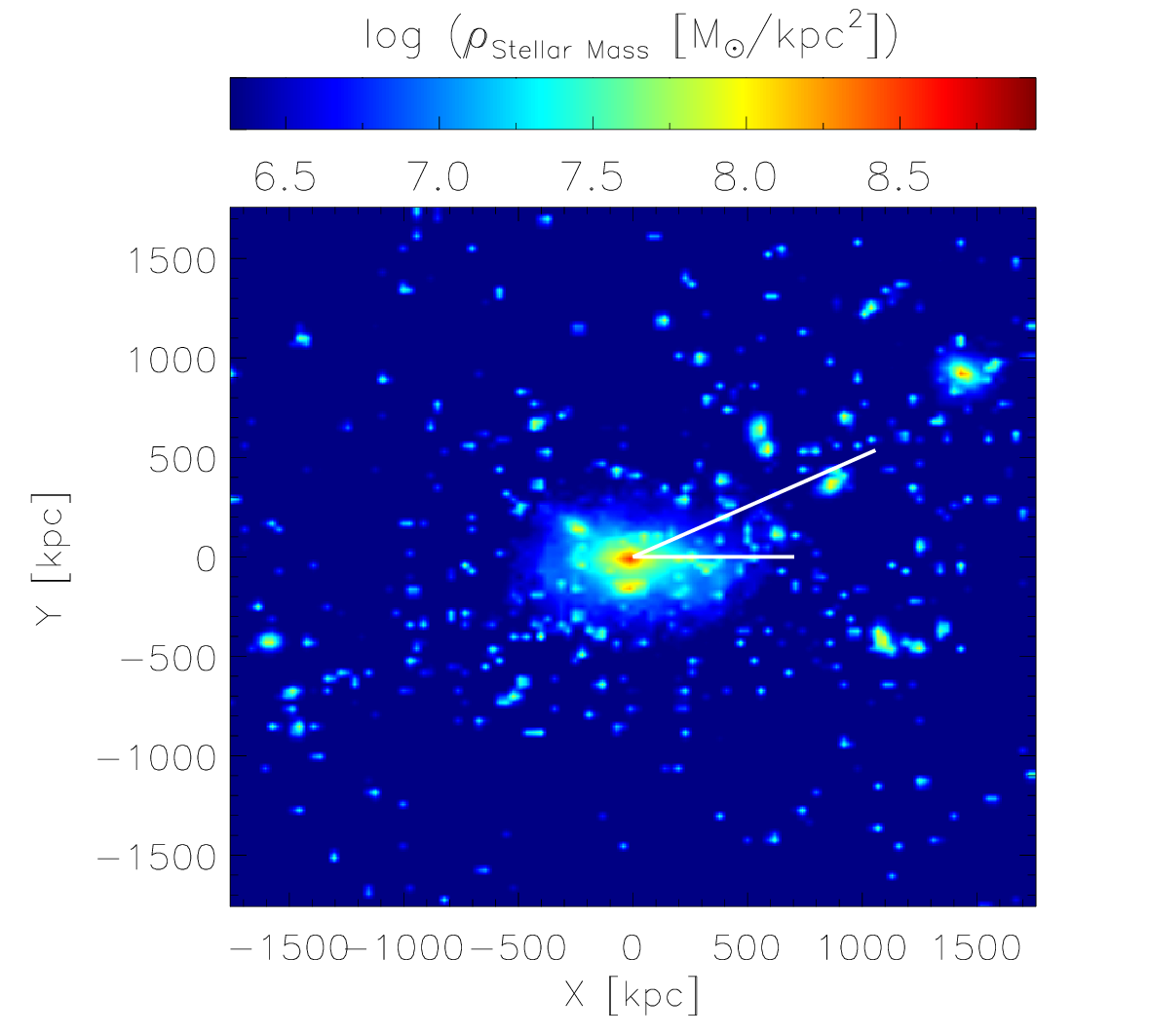}
    \includegraphics[width=5.5cm]{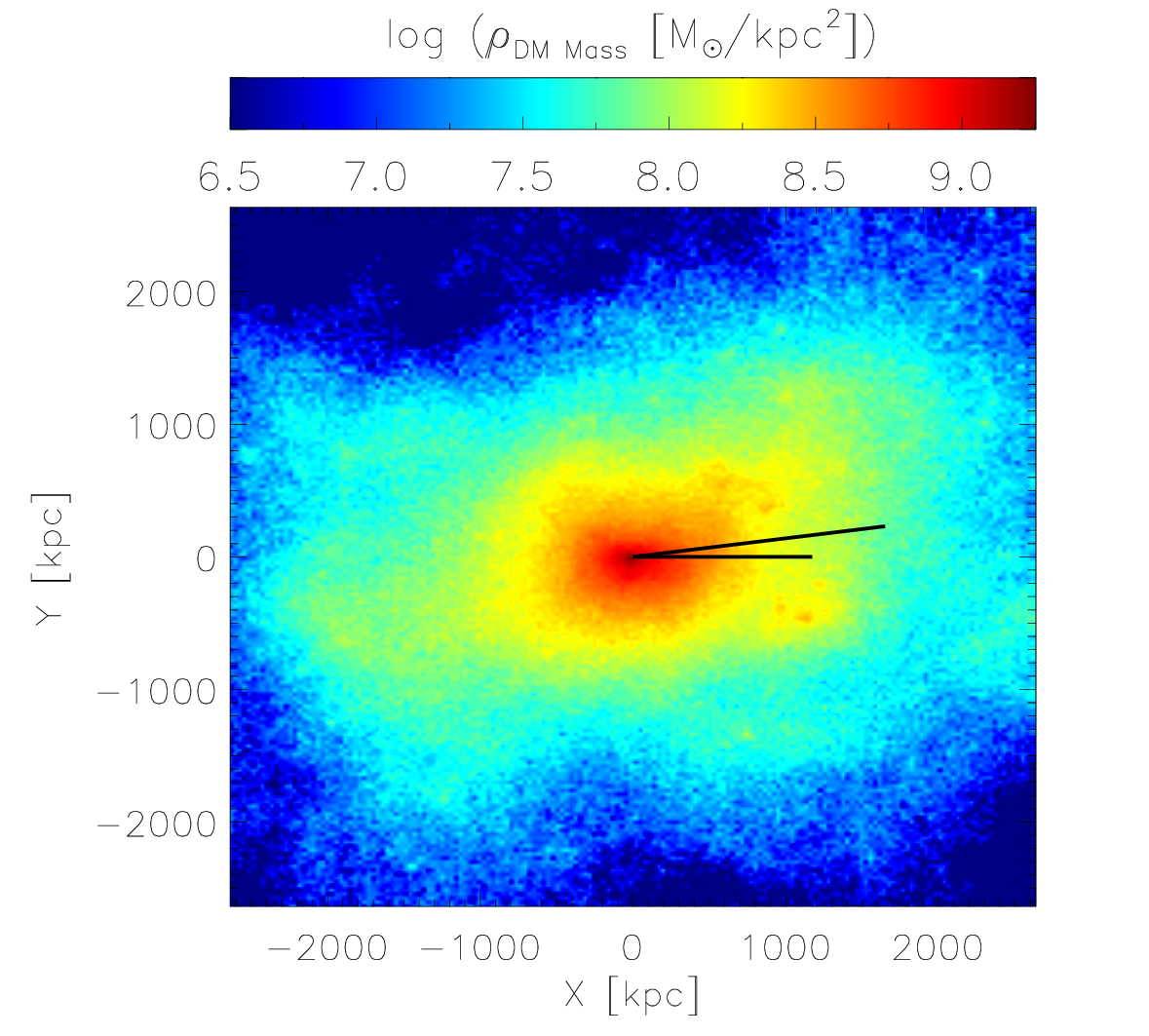}
    \caption{An example of the different mass distributions and their elongation axis for one of our simulated clusters at $z=0$. Left Panel: BCG stellar mass distribution. The short line corresponds to the BCG elongation axis, this is set to coincide with the x-axis in all panels. The long line marks instead the cluster galaxies elongation axis. Middle Panel: satellite galaxies distribution, short and long lines are as in the left panel. Right panel: cluster DM distribution. The long line follows the direction of the cluster DM elongation axis.}
    \label{fig:maps}
\end{figure*}

\section{Methods}
\label{sec:method}

Cluster of galaxies are identified by means of a friends-of-friends (FOF) algorithm. First we link DM particles in the high resolution regions with a linking length of 0.16 times the mean inter-particle separation. Gas and star particles are then linked to the FOF group defined by the DM particles using the same linking length.

Galaxies inside the clusters are instead identified using the {\footnotesize SUBFIND} subhalo finder algorithm \citep{springel2001, dolag2009}. This algorithm uses all the particles in the FOF group to determine saddle points of the density field, and then groups together all those particles lying inside a border defined by the spatial position of such saddle points. Then, an unbinding procedure is applied to eliminate high speed particles (i.e., those not gravitationally bound to the substructure). The unbound particles are assigned to the main subhalo. The latter  includes all particles not belonging to any other subhalo, thus also the BCG and the intra-cluster stars.

%\CIN{Giuseppe, I am using SPOS. I thought that SPOS was the position of the minimum potential of the subhalo, but instead there is a block containing the most bound particle id of subhaloes 'MBID'. The problem is that the position of this most bound particle does not coincide exactly with spos. I would need to know What exactly SPOS and GPOS are}.
The center of the halo and the BCG coincide and is given by the particle belonging to the main SubFind subhalo having the minimum value of the gravitational potential. This center is then used to compute at each redshift $R_{200}$ and $R_{500}$ along with their associated masses $M_{200}$ and $M_{500}$, respectively.

In all figures the BCG is defined as the stellar particles that belong to the main cluster subhalo and are within 10 per cent of $R_{500}$ of the center. Nevertheless, for sake of comparison we also consider BCG as particles inside 50 kpc (physical kpc).
As shown in \cite{ragone2018} the 10 per cent of $R_{500}$ at $z=0$ is similar to the radius at which our simulated BCGs drop to a rest-frame surface brightness of $\mu_V \sim 24 {\rm~mag~arcsec^{-2}}$, a classical observational value to define the galaxy limit \citep{devaucou1991}.
%In our sample of clusters, $0.1R_{500}$ amounts on average to $\sim$ 155 pkpc, while  $\mu_V \sim 24 {\rm~mag~arcsec^{-2}}$ is reached at $\sim$ 130 pkpc.
In our sample of clusters, $0.1R_{500}$ amounts on average to 155, 55 and 30 kpc at redshift 0, 1 and 2, respectively \citep[see Fig. 7 in][]{ragone2018}.
%, while  $\mu_V \sim 24 {\rm~mag~arcsec^{-2}}$ is reached at 130, 110 and 100 kpc at redshift 0, 1 and 2, respectively.

The alignment between our simulated BCGs and their host clusters can be quantified by means of the angle between the elongation axes of both stuctures.
These elongation axes can be obtained from the principal axes of the ellipsoids that best describe the corresponding distribution of matter. The common practice is to obtain these principal axes from the eigenvectors of a shape tensor that can be expressed as
\begin{equation}
\mathcal{S}_{ij}=\frac{1}{M}\sum_n{m_n w_n x_{n,i}~x_{n,j}}
\label{eq:shapetens}
\end{equation}
where $x_{n,i}$ and $x_{n,j}$ are the $i^{\rm th}$  and $j^{\rm th}$ component of the $n^{\rm th}$ particle position vector relative to the system center, $m_n$ is its mass, $M$ is the sum of the $m_n$ masses and $w_n$ is a weight that is typically related to the distance of the particle to the system center.

The length of the ellipsoid semi-axes \((a>b>c)\) are related to the eigenvalues, whereas the directions of the corresponding principal axes \((\hat{a}, \hat{b}, \hat{c})\) are provided by the eigenvectors of the shape matrix.
These computations can be performed iteratively, as proposed for example by \citet{zemp2011}. The iterative technique consists in repeating the determination of the ellipsoid until some convergence criteria is satisfied. We first compute $\mathcal{S}$ using all the particles, in the BCG or in the cluster (inside $\rtwo$), yielding initial $a,b$ and $c$.
%Next the lengths of the principal axes of the ellipsoid are rescaled keeping the enclosed volume constant.
New $a,b$ and $c$ are next determined discarding particles outside the initial ellipsoidal volume. The process is repeated until changes in the axis ratio become smaller than 0.001. However when dealing with observations, iteration is not used.

For our simulated clusters we compute the best ellipsoids in two ways: using the galaxies (ClusterGlxs) or using the DM  particles (clusterDM) inside $\rtwo$. We consider as galaxies subhaloes with stellar masses $> 1 \times 10^{10} \msun$.

When the cluster ellipsoid is estimated with the galaxies we consider four cases in Eq. \ref{eq:shapetens}:
\begin{itemize}
\vspace{-\topsep}
\item \mw:  $m_n$ is the mass of the galaxies and $w_n=1$
\item \rw: $w_n$ is the inverse of the square distance of galaxies to the cluster center and $m_n=1$
\item \mrw: $m_n$ is the mass of the galaxies and $w_n$ is the inverse of the square distance of galaxies to the cluster center
\item \now: $m_n=1$ and $w_n=1$
 \end{itemize}
\vspace{-\topsep}
Since we apply both the iterative and non-iterative computations of the best ellipsoids, we obtain eight estimates of the cluster galaxy distribution principal axes.

Regarding the DM halo we follow \citet{zemp2011} and use the iterative technique without weights removing DM subhaloes before any computation. This leaves us with only one estimation of the best ellipsoid for the DM halo.

The BCGs best ellipsoids are computed using both the iterative and non-iterative techniques, applied only to star particles, for the \mw\ (with $m_n$ equal to the star particles mass), \rw, \mrw\ and \now\ cases. We hence obtain eight estimations of the BCG principal axes.

Then, in the 3D case, for each BCG-Cluster pair we compute nine alignment angles, one between the central galaxy (iterative \mw) and the cluster DM halo (iterative \now) and the remaining eight between the central galaxy and the cluster galaxies (4 iterative and 4 non-iterative computations of the \mw, \rw, \mrw\ and \now\ cases).

The 3D alignment angle $\alpha$ is defined as the acute angle between the principal major axes of the BCG and the cluster
\begin{equation}
\alpha={\rm arccos}(|\hat{a}_{BCG} ~\cdotp~ \hat{a}_{Clus}|)
\label{eq:alpha}
\end{equation}
If these two major axes were randomly oriented then our sample of clusters would have a median $\alpha = 60^\circ$ \footnote{We found some confusion in the literature on this point. For a random orientation in 3D the median angle is $60^\circ$, while the mean angle is $\simeq 57.3^\circ$  (1 radian). In some works the distribution of $\cos \alpha$ is considered, whose mean is 0.5 which corresponds to $\simeq 60^\circ$}, and a 25\%-75\% percentiles of $\sim 41.4^\circ$ $\sim 75.5^\circ$ respectively.

The projected shapes are computed considering the three orthogonal lines of sight. Only the non-iterative \now~ computation is used here, but in order to mimic what it is done in observations we also consider a case were only the 20 most massive galaxies are used to obtain the shape and position angle of clusters.
The projected BCG-Cluster alignment is measured in two ways:
\begin{itemize}
\vspace{-\topsep}
\item $\alpha$ alignment, obtained from Eq. \ref{eq:alpha} but using the elongation axis of the projected BCG and cluster mass distributions. The mean (and median) angle expected for a uniform random distribution of projected orientation, that is in absence of any alignment signal, is $45^\circ$ with a standard deviation of $\sigma=90^\circ/\sqrt{12}=25.98^\circ$, and 25\%-75\% percentiles of $22.5^\circ$  and $67.5^\circ$.

\item $\theta$ alignment \citep[e.g.][]{yang2006,hao2011}, obtained from
\begin{equation}
\theta = \frac{1}{N}\sum_{n=1}^{N}\theta_n
\label{eq:theta}
\end{equation}
where $\theta_n$ is the angle between the projected BCG major axis and the line connecting the BCG to the projected position of the $n^{\rm th}$ satellite galaxy. If the BCG preferentially aligns with the distribution of cluster satellite galaxies, then it should be obtained $\theta < 45^\circ$.
In a given sample of clusters, if the BCG principal axes are randomly oriented with respect to the cluster satellite distributions, then $\langle\theta\rangle = 45^\circ$ is expected. The computation of the corresponding standard deviation is not straightforward since it depends on the angular distribution of galaxies inside each cluster. In order to cope with this, we compute the standard deviation numerically (at every simulation output) after one random shuffling of the BCG elongation axis in each cluster.\footnote{If instead of shuffling the BCG elongation axis we randomly shuffle the satellite galaxies angular positions ($\theta_n$ in Eq. \ref{eq:theta}), then $\langle\theta\rangle = 45^\circ$ and $\sigma = 90^\circ/\sqrt{12 N}$, where $N$ is the number of galaxies used to compute $\theta$, provided it is the same for all clusters have. This latter operation samples the BCG $\theta$ alignment with a uniform angular distribution of galaxies. Since real clusters are triaxial systems, this is not a correct representation of the random distribution of $\theta$.}
\end{itemize}

An example of the different matter distributions for one of our simulated clusters at $z=0$ can be seen in  Fig. \ref{fig:maps}. Left, middle and right panels depict the BCG stars, cluster galaxies and cluster DM components, along with the corresponding elongation axes.

In order to reconstruct the evolutionary path of each cluster, we follow back in time its main progenitor from $z=0$ to $z=4$ using 71 simulation outputs. We compute at each redshift the best ellipsoid of both the cluster and its central galaxy.

\begin{figure*}
    \includegraphics[width=\columnwidth]{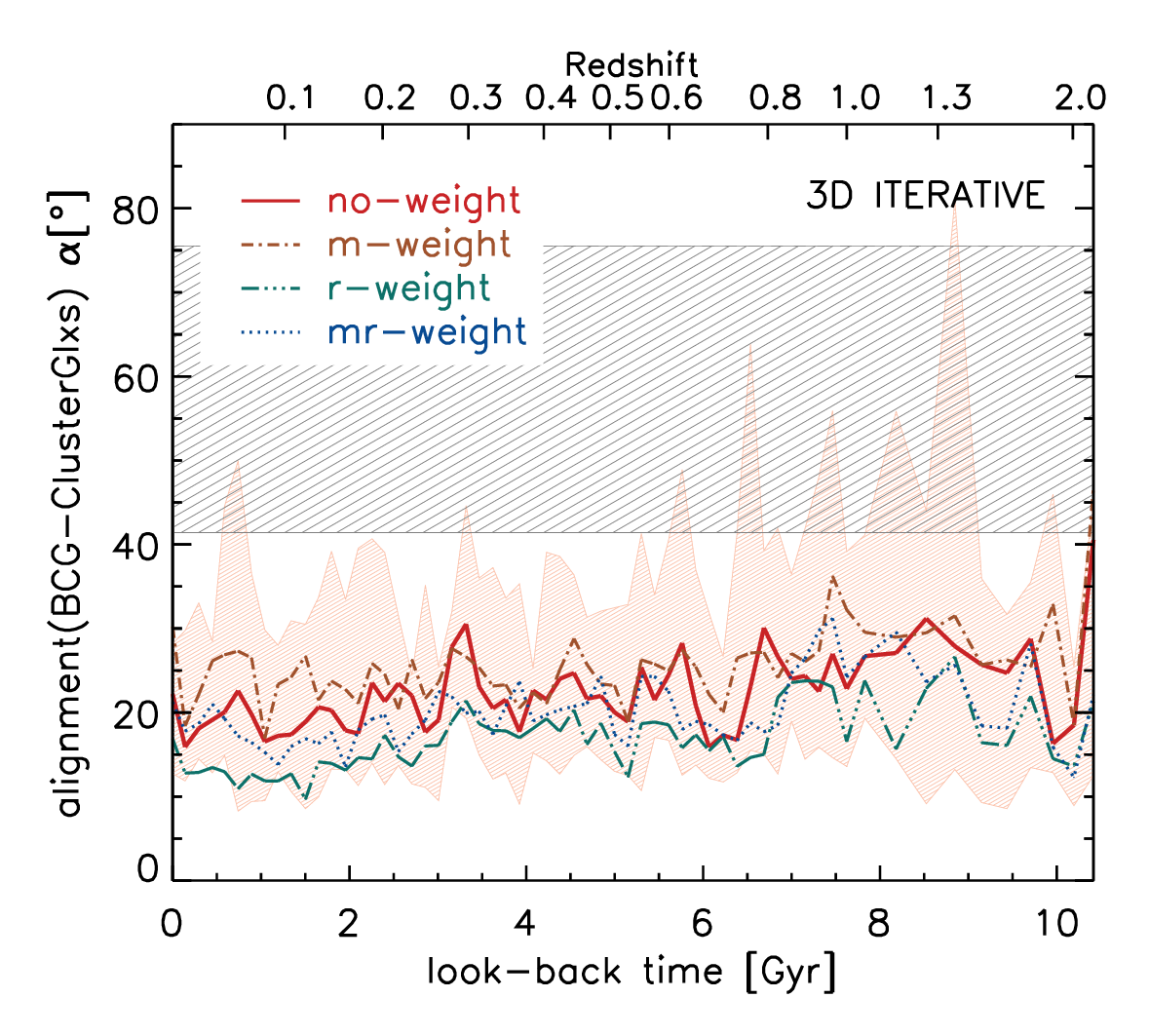}
    \includegraphics[width=\columnwidth]{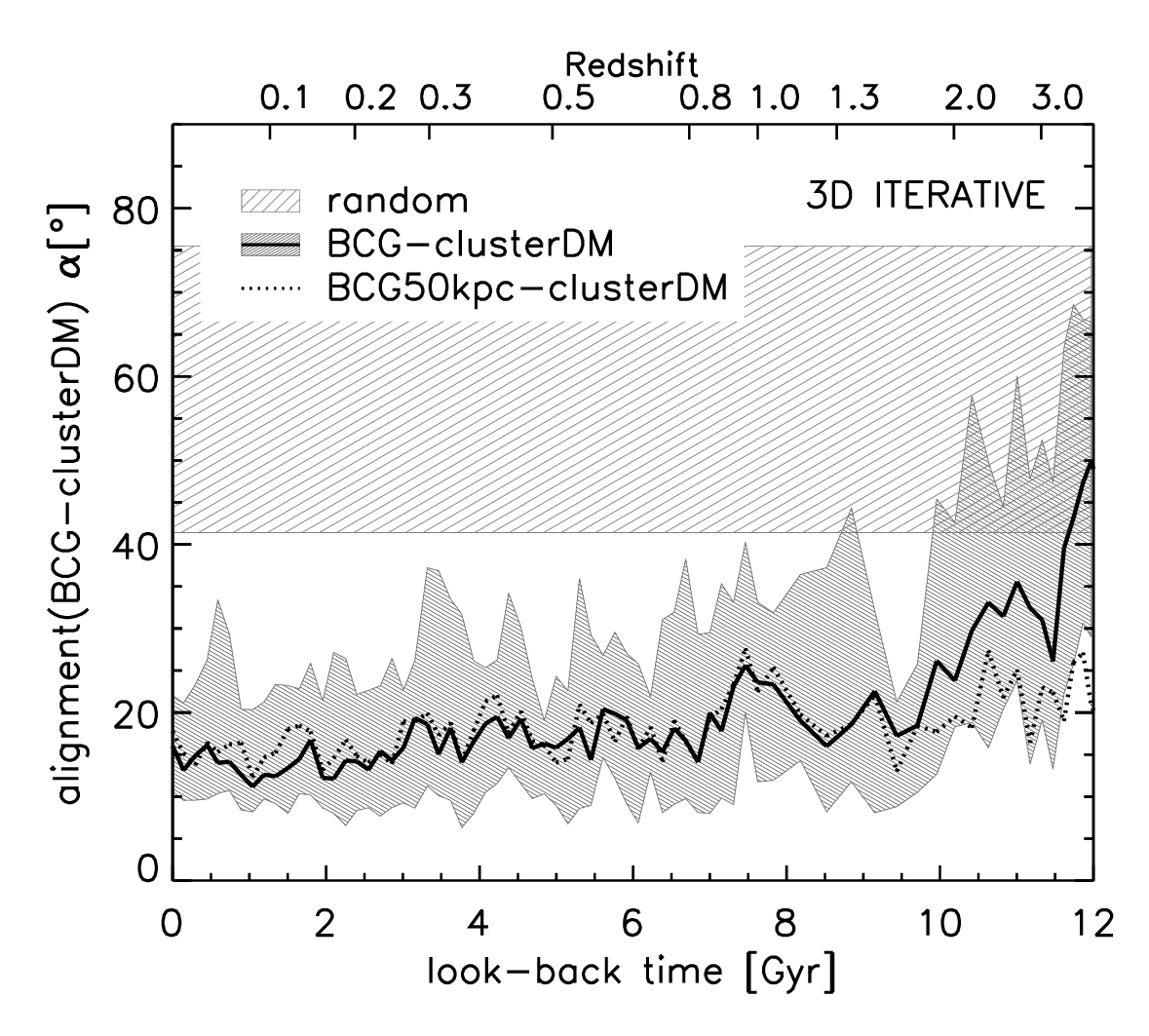}
    \caption{Median values of BCG-ClusterGlxs (left panel) and BCG-ClusterDM (right panel) alignment angles for our set of 29 simulated cluster at each epoch (note that limits in the x-axis are different in both panels, see text). The shaded area encloses the 25\%-75\% percentiles of the distributions. For sake of clarity in the left panel only the \now~ case percentiles are shown.
    Up to $z\lesssim2$ there is no clear evolution of the alignment. At $z\gtrsim2$ the BCG-ClusterDM alignment angle increases gradually toward earlier epochs, nevertheless this is not the case if BCG is defined as stars particles inside a 50kpc fixed aperture (dotted line). The horizontal dashed area corresponds to the 25\%-75\% percentiles of the angle distribution for random directions in 3D (see text).}
    \label{fig:alig_evo}
\end{figure*}

\begin{figure*}
    \includegraphics[width=\columnwidth]{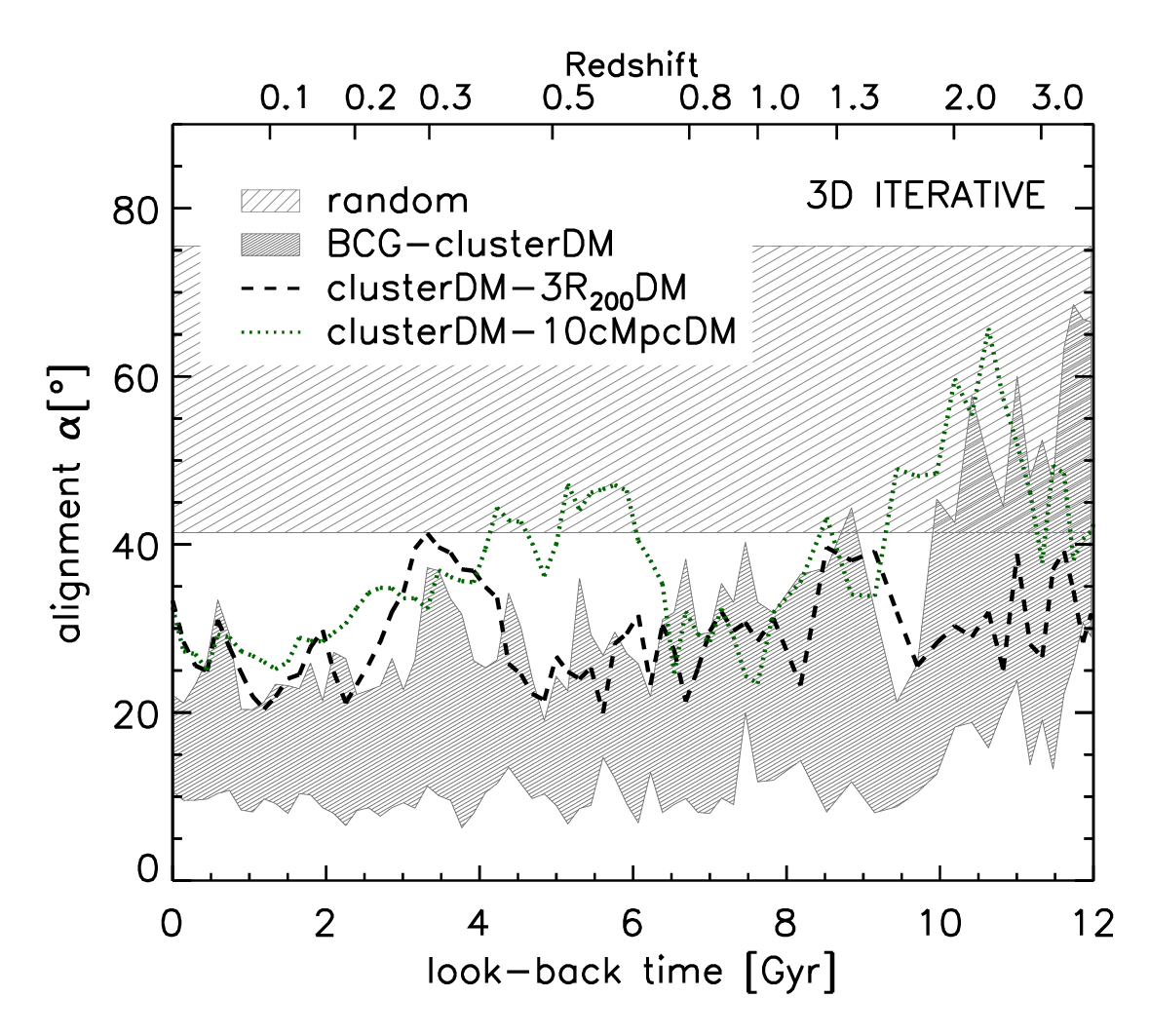}
    \includegraphics[width=\columnwidth]{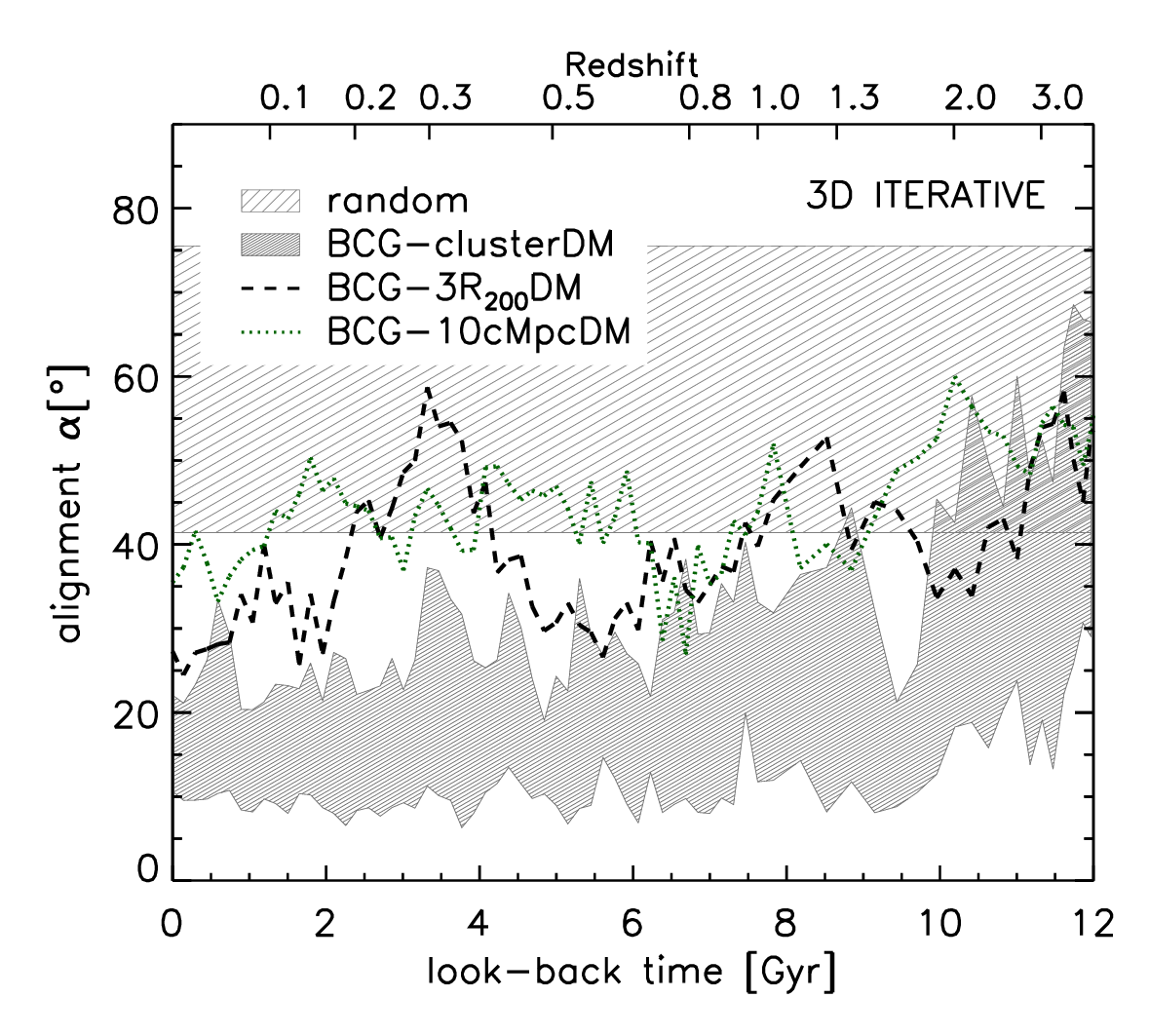}
    \caption{In the two panels, to ease the comparison we include again the shaded area that encloses the 25\%-75\% percentiles of the BCG-clusterDM alignment distributions (as in right panel of Fig. \ref{fig:alig_evo}). The dashed area shows instead the 25\%-75\% percentiles of the alignment angle distribution for random orientations. Left panel: Dashed and dotted lines stand for the medians of the alignment between the clusterDM and the DM within 3$R_{200}$ and 10cMpc respectively.
    During the time interval studied in this work clusters have always been aligned with their nearby surroundings (3$R_{200}$) with an angle $\alpha \lesssim 30 ^\circ$ (dashed line). The alignment with the 10cMpc scale is somewhat weaker and seems to develop later at $z\sim 1.5$. The median of 3$R_{200}(z=0)\sim$7cMpc and that of 3$R_{200}(z=1)\sim$5.5cMpc.
    Right panel: Dashed and dotted lines stand for the medians of the alignment between the BCG and the DM within 3$R_{200}$ and 10cMpc respectively. Though to a lesser extent than ClusterDM, the BCG is also aligned with the larger scale distribution of mass. The strength of this alignment weakens when computed with the mass at larger scales and at higher redshifts.}
    \label{fig:lss}
\end{figure*}

\begin{figure}
    \includegraphics[width=\columnwidth]{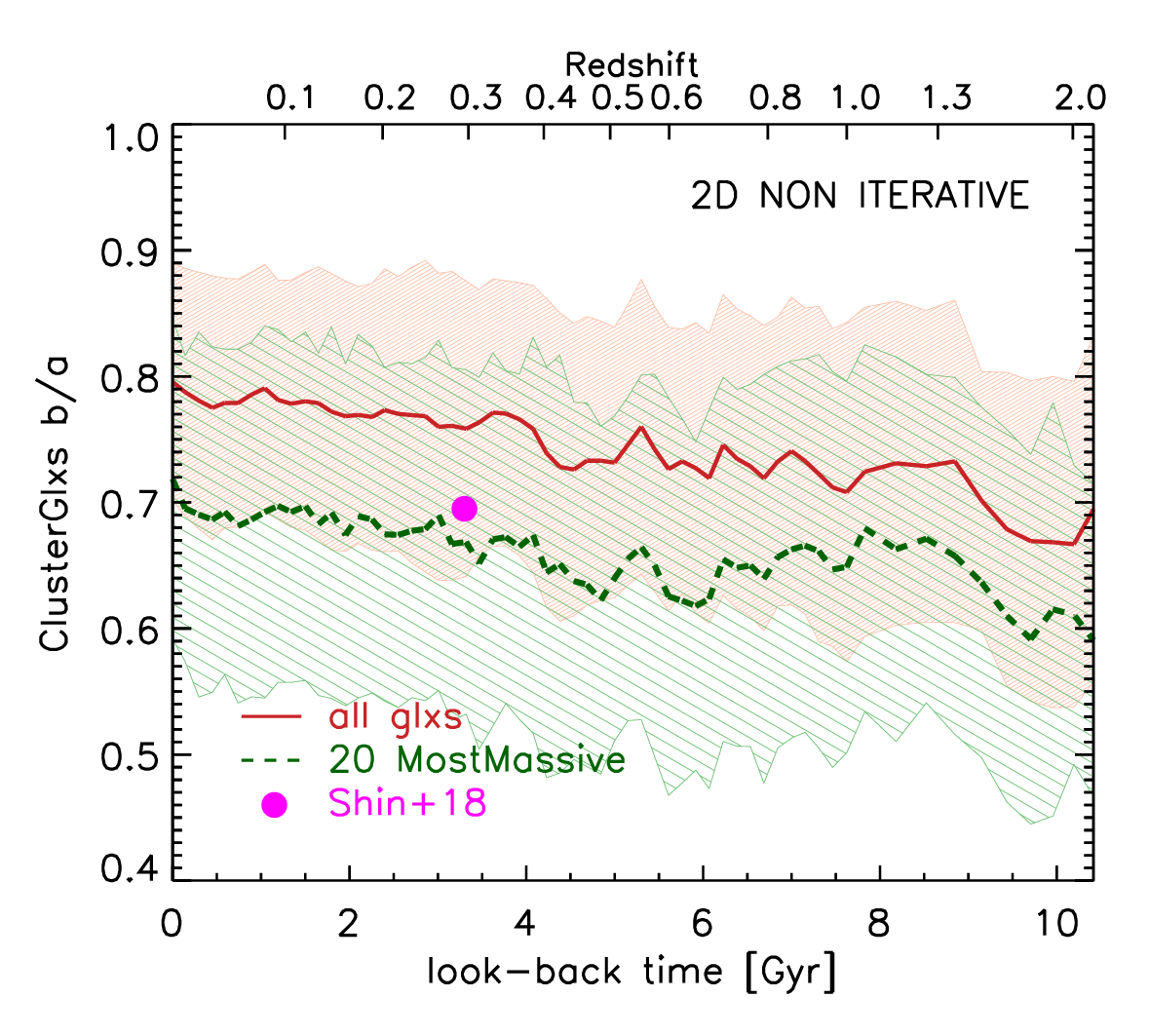}
    \caption{The evolution of the simulated cluster sample mean projected shape $b/a$ with time and the effect of {\it noise} bias. Clusters mean $b/a$ measured with all cluster galaxies (solid line) is larger than $b/a$ computed with only the 20 most massive galaxies (dashed line). In both cases clusters seem to have more rounded projected shapes at lower redshift. The shaded areas encloses the 25\%-75\% percentiles of the $(b/a)$ distribution at each simulation output. The large dot at $z\sim0.3$ corresponds to a recent observational estimation \citep{shin2018} before being corrected by noise bias, see text.}
    \label{fig:shape_evo_2D}
\end{figure}

\begin{figure*}
    \includegraphics[width=\columnwidth]{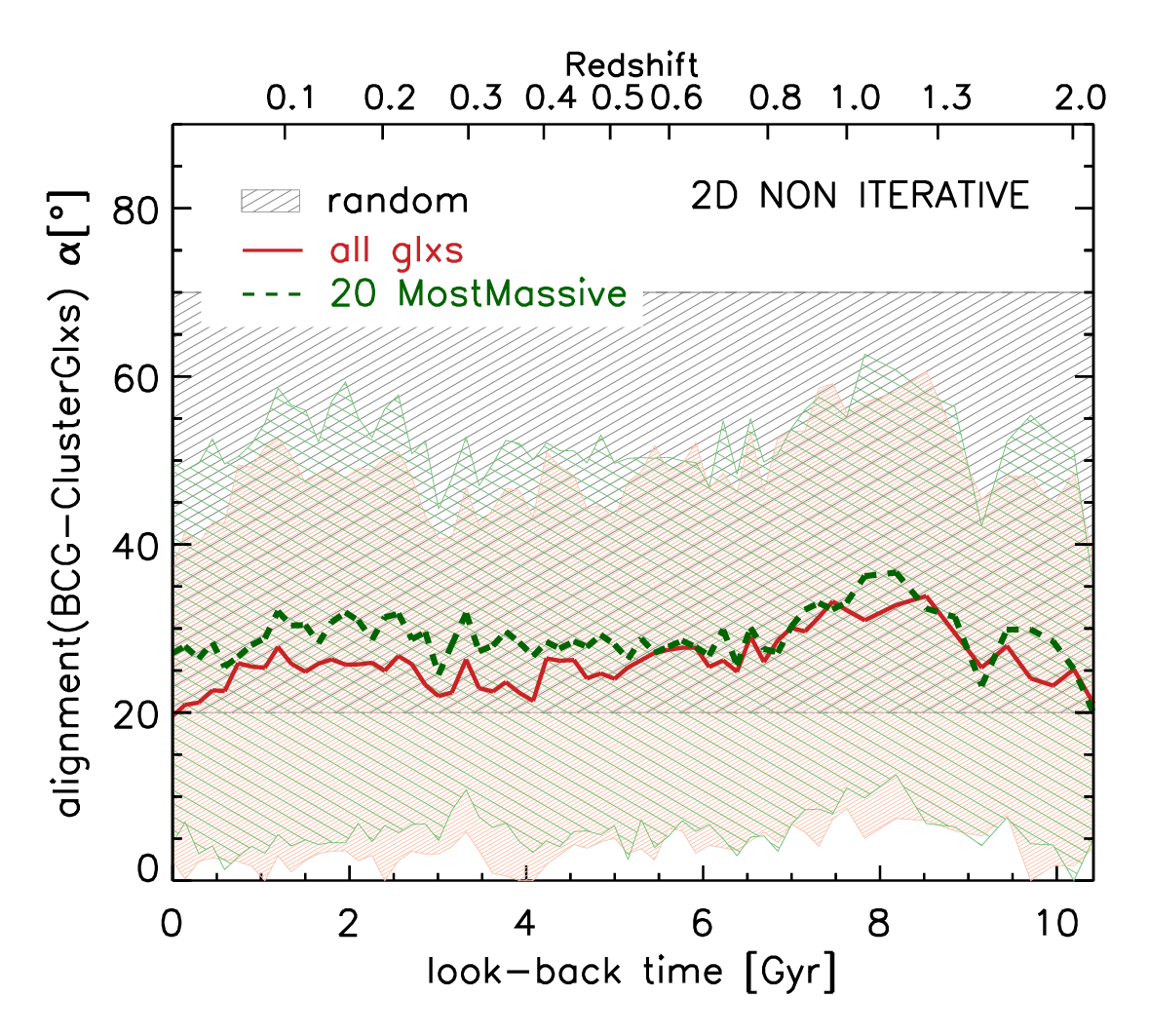}
    \includegraphics[width=\columnwidth]{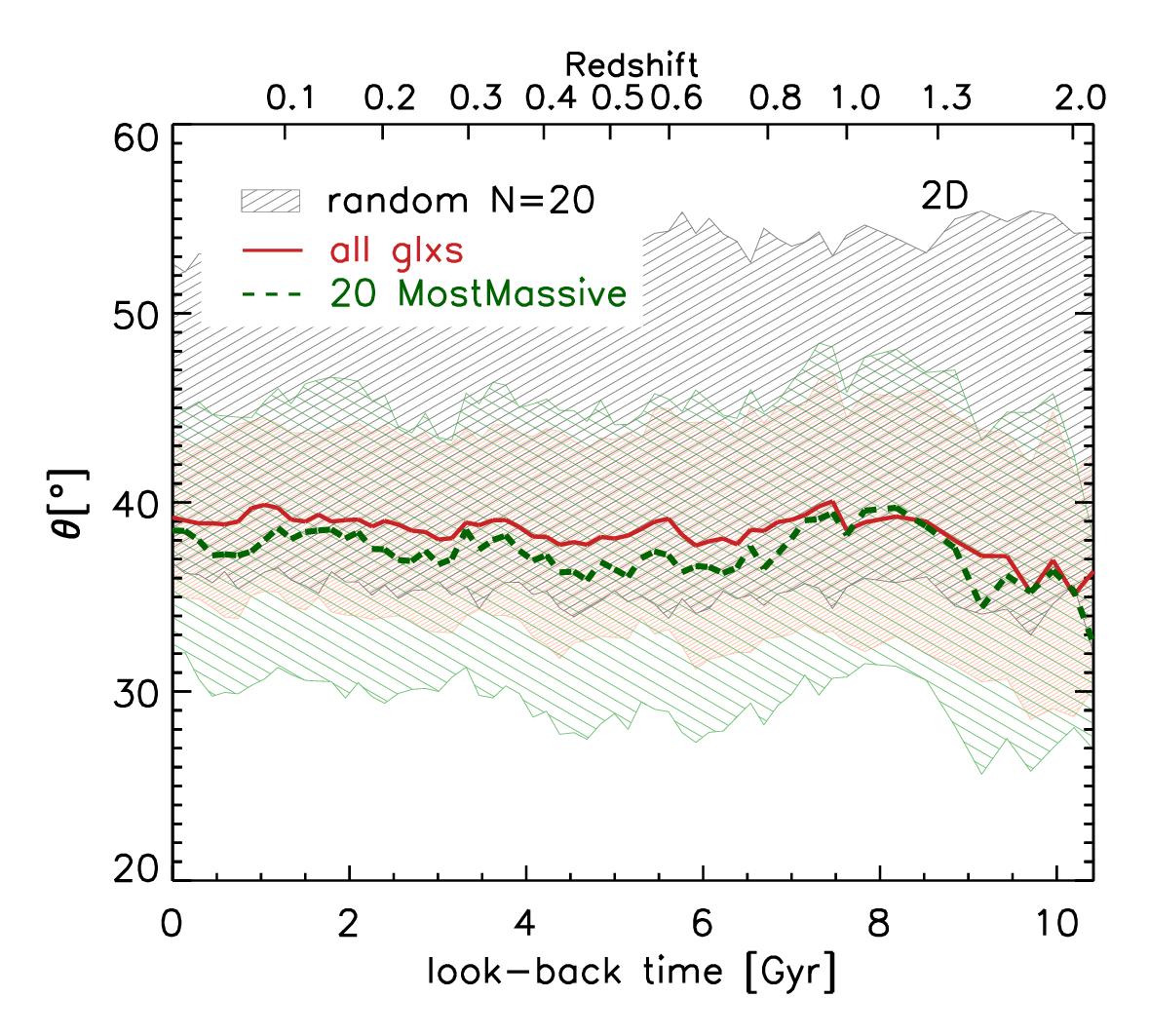}
    \caption{The mean projected $\alpha$ (left) and $\theta$ (right) alignments considering the 3 possible projections for the sample of 29 simulated clusters. Computations using all (solid lines) and the 20 most massive galaxies (dashed lines) are shown. Shaded areas surrounding each mean correspond to the $\pm1\sigma$ of the corresponding distributions.
    Left panel: The projected $\alpha$ alignment is somewhat looser than the full 3D alignment but is still present in the whole studied redshift range. The horizontal dashed area with center at $45^\circ$ encloses the $\pm1\sigma$ deviation of the random alignment expectation.
    Right panel: The projected $\theta$ alignment is also present during the whole redshift range. The dashed area surrounding $45 ^\circ$ corresponds to the $\pm1\sigma$ standard deviation of a random distribution of $\theta$, which is computed after one shuffling of the BCG elongation axis in each cluster (see text).}
    \label{fig:alig_evo_2D}
\end{figure*}

\section{Results}
\label{sec:result}
\subsection{3D Alignment}
We start by showing in the left panel of Fig. \ref{fig:alig_evo} the evolution of the median alignment angle ($\alpha$) between the principal axes of BCGs and their host clusters (BCG-ClusterGlxs alignment), where the cluster shapes and principal axes have been computed using galaxies. For sake of brevity we concentrate in the iterative technique only, but we verified that conclusions hold true when using the non iterative computations.
The maximum look-back time in this panel is given by the condition that cluster ellipsoids must be computed with at least 20 galaxies.

As found in observations, we obtain a very clear signal of BCG-ClusterGlxs alignment at z=0. The distribution of the alignment angles is tight, for instance in the \now~ case we find a median of 22.2$^\circ$ and 25\% and 75\% quartiles of 12.7$^\circ$ and 28.2$^\circ$ respectively.
The alignment signal for all the BCG-ClusterGlxs pairs of ellipsoids (\mw, \rw, \mrw and \now) persists over the whole considered redshift range, with a very mild tendency to increase with time. The strength of the alignment signal can be appreciated by comparing with the horizontal dashed area in the figure. The latter covers the same percentiles range than before, but for a distribution of angles between two randomly oriented axes.

%In fact, the strength of the alignment signal can be visualized by comparing with the dashed area which encloses the 25\%-75\% percentiles of alignment angles in a control sample which is created as follows. We keep fixed the BCG major axis and compute its alignment with the cluster principal axis where the position angles of galaxies have been randomly shuffled by an isotropic angle distribution. The galaxy distances to the center of the cluster is not modified. For each cluster we compute the mean random alignment angle after performing 1000 realizations of this shuffling.

The right panel of Fig. \ref{fig:alig_evo} depicts the evolution of the alignment angle between the BCG and the DM halo (BCG-ClusterDM alignment) as a function of look-back time. The alignment obtained with the usual definition of BCG is shown with the solid line, whereas the dotted line is for the alignment with the BCG defined as stellar particles inside 50kpc. At $z\gtrsim 1.5$ this fixed aperture is typically greater than the fiducial 10\%$R_{500}$ aperture and can progressively include a significant fraction of the cluster main progenitor. Hence, the determination of the main galaxy position angle might be affected by the distribution of stellar matter outside the galaxy, which could artificially increase the alignment signal.
Nevertheless, during the last 10Gyr both BCG-ClusterDM alignments (with BCG defined as star particles inisde both 50kpc or 10\%$R_{500}$) are systematically stronger than the BCG-ClusterGlxs one (left panel), with a median usually below $20^\circ$. Once again, we find just a very weak, if not negligible, tendency for a better alignment with time. By converse, \cite{okabe2019} claim for a clear improvement of the alignment toward $z=0$. However, we note that they study a significantly less massive set of clusters and do not remove subhaloes to describe the DM distribution.

%analyzed the evolution of the BCG-ClusterDM alignment for the 40 most massive haloes ($M_{200} \gtrsim 5 \times 10^{13}\, \msun$) in a 100 $h^{-1}$cMpc box hydro-simulation.
% ellos <cos>=0.82 a z0   0.70 a z0.7
%  nos  <cos>=0.9

%Nevertheless, at $z\gtrsim2$ the curves exhibit different behaviours. Namely, the BCG-ClusterDM alignment continues to be present only if the 50kpc aperture is used to define BCG. Otherwise, it worsens gradually toward earlier epochs where it approaches the random alignment distribution signal (horizontal dashed area).
%

On the other hand, the worsening of the BCG-ClusterDM  alignment at earlier times ($z \gtrsim 2$) could be simply related to the fact that at those redshifts the central galaxy is ill defined. Inside the proto-clusters there is not an obvious dominant galaxy but instead several galaxies which compete in mass.
Another important fact to consider is that interactions and mergers are more frequent at early time.
%However, this is an aspect that needs to be studied in more details since there has been recent claims suggesting that in the local Universe the BCG-Cluster alignment would be robust against cluster major mergers \citep{wittman2019}.
We will return to this point in Section \ref{subsec:mergers}.

In order to further analyze the evolution of the alignment with larger scales Fig. \ref{fig:lss} depicts the alignment of cluster DM (left panel) or BCGs (right panel) with the distribution of matter within 3$R_{200}$ (dashed line) and 10cMpc (dotted line). To ease the comparison we include the 25\%-75\% percentiles shaded area of the BCG-ClusterDM alignment presented in the right panel of Fig. \ref{fig:alig_evo}.
It is interesting to note that the alignment of the cluster DM halo with the distribution of matter within 3$R_{200}$ (median 3$R_{200}|_{z=0}\sim$7cMpc, median 3$R_{200}|_{z=1}\sim$5.5cMpc) is present over the whole studied redshift range whilst that with the even larger scale of 10cMpc begins to be clearly distinguishable from random alignments only at $z\lesssim1.5$.
A 10cMpc scale seem to be exceedingly large at $z\gtrsim1.5$ as to be correlated with the protocluster.
Though to a lesser extent than ClusterDM, BCGs show also indications of being aligned with the larger scale field. The strength of this alignment weakens when computed with the mass at larger scales or at higher redshifts.
This could imply that the alignment stem from the outside, correlating first a larger scale with the cluster and then the cluster with the central galaxy.

\begin{figure}
    \includegraphics[width=\columnwidth]{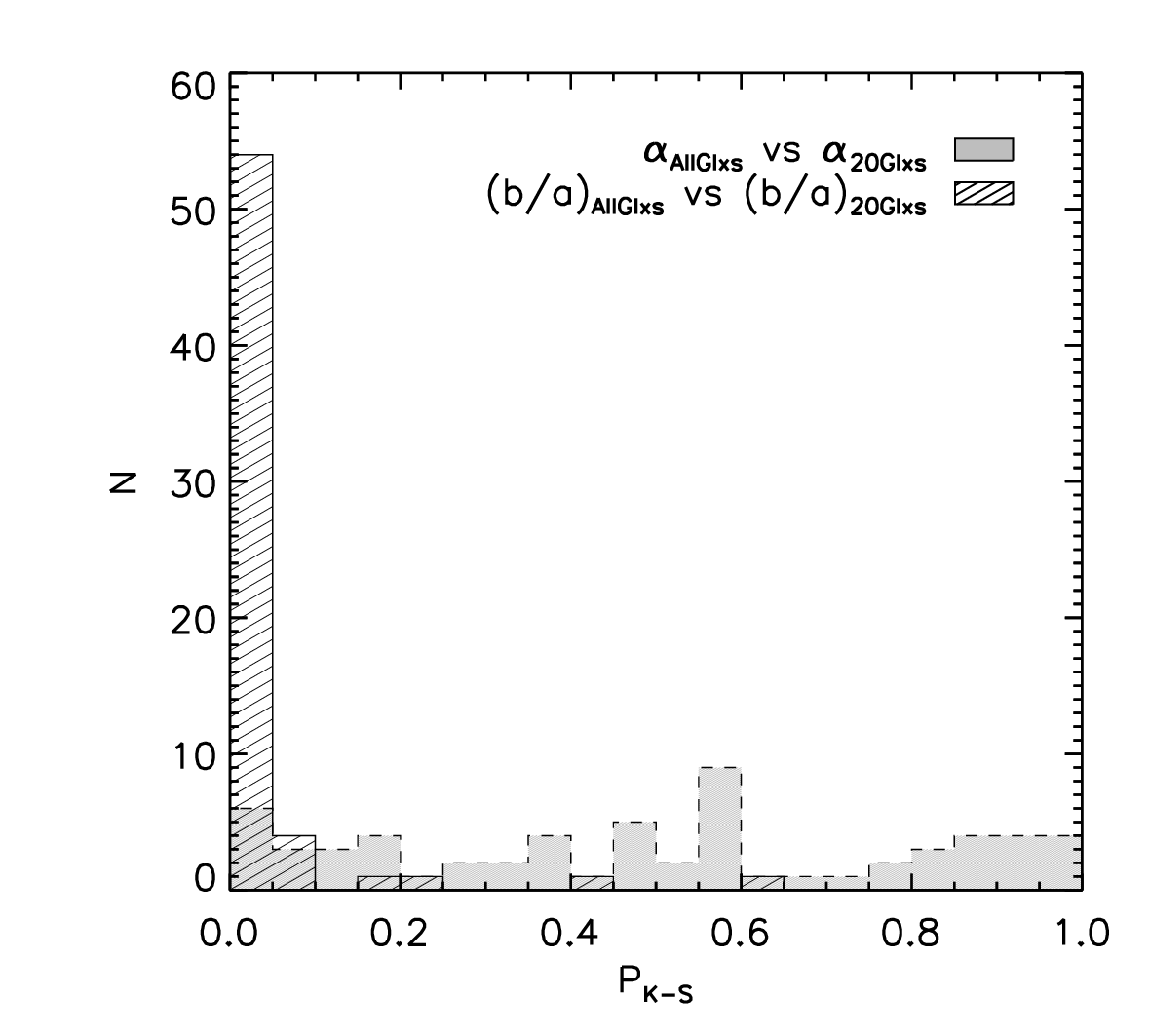}
    \caption{Effect of the {\it noise} bias on clusters $b/a$ and the projected BCG-Cluster alignment $\alpha$.
    $P_{K-S}$ is the Kolmogorov-Smirnov probability that two samples are drawn from the same distribution.
    The dashed histogram shows the $P_{K-S}$ values obtained when comparing the two $b/a$ distributions in Fig. \ref{fig:shape_evo_2D} at each redshift. $P_{K-S}$ has a sharp peak at small values, implying that clusters $b/a$  are substantially different if computed with all or with only 20 galaxies.
    On the contrary, the position angles of clusters seem to be less affected by this so-called {\it noise} bias. In fact, the gray solid histogram of $P_{K-S}$ values resulting from the comparison of the two $\alpha$ distributions in the left panel of Fig. \ref{fig:alig_evo_2D}, are comparatively large.}
    \label{fig:pks}
\end{figure}

\subsection{Projected Shape and Alignment}
In order to more closely compare with observational results we compute the projected cluster shapes and the BCG-ClusterGlxs alignment between the projected distributions of BCG stars and cluster galaxies.
As mentioned before we consider three possible projections.
The evolution in time of the mean\footnote{In this section, whose results are comparable with observation, we plot the mean instead of the median. Although the latter is generally a more useful statistic of the distribution, usually in observational works the former is considered.} minor-to-major axis ratio $b/a$, of the projected galaxy distribution is shown with solid line in Fig. \ref{fig:shape_evo_2D}.
There is a mild evolution of $b/a$ indicating that clusters evolve toward rounder shapes at lower redshifts.
This evolution might be partly due to a well known artifact, dubbed {\it noise} bias, created by discreteness. This artificially increases ellipticities with decreasing sampling \citep{paz2006,plionis2006,ragone2010,shin2018} and in fact  this could well be the case since clusters at higher redshifts are progressively populated with less galaxies.
In order to better understand how this artifact affects our clusters we recompute axis ratios using always a fixed number of galaxies picking up the 20 most massive ones. This choice mimics somehow the magnitude limit that is present in observational catalogues eliminating in turn the sampling number effects.
As expected when working with a lower number of objects we obtain lower values of  $b/a$ than when using all the available galaxies, but the mild correlation of the median shape with time remains. Moreover, $(b/a)$ at low redshifts are very similar to the mean value reported by \cite{shin2018} for their most massive clusters, before they correct for {\it noise} and {\it edge} bias (see their Fig. 2).

As mentioned in Section \ref{sec:method} we have two estimates of the projected alignment, namely the projected version of Eq.\ref{eq:alpha} and the mean angle between the BCG elongation and the distribution of the satellite galaxies defined in Eq.\ref{eq:theta}. Fig. \ref{fig:alig_evo_2D} shows the $\alpha$ and $\theta$ alignment in the left and right panel respectively compared with the corresponding random expectations. In both panels, it can be seen that the BCG-Cluster alignment is present also in 2D projections up to redshift $z \lesssim 2$ with no clear signs of evolution. The same is true if the sample of clusters with richness equal to 20 galaxies is used.  However, the projected $\alpha$ alignment signal is somewhat less evident than in the 3D case.

On the observational side the evolution of the alignment with time is far from being well assessed. On one hand, \citet{huang2016} analyzed the alignment phenomenon in a sample of 8237 clusters constructed from the Sloan Digital Sky Survey in the redshift range 0.1 to 0.35 and with estimated masses $M_{200}\gtrsim 1.4 \times 10^{14} M_\odot$. They reported an average difference in the position angle of the BCG and the cluster of 35$^\circ$, with no evidence of redshift dependence in their limited range, in agreement with our result.
On the contrary, using cluster samples up to redshift $\lesssim$0.44 it has been found a stronger alignment signal as redshift decreases (\citealt[][]{niederste-ostholt2010}; \citealt[][]{hao2011}).

%\CIN{en el trabajo de \citet{zhang2019}, tita 2D en simulaciones, encuentran mayor alineamiento a z alto. Usan z=0,1,2,3,5. Pero no se entiende como esta compuesta la muestra de cumulos}

Surprisingly, and contrary to the $b/a$ of clusters, at any redshift the $\alpha$ alignment angle distribution we obtain with the 20 most massive galaxies is very similar to the distribution derived by using all the galaxies inside the cluster. This can be  quantified by means of the Kolmogorov-Smirnov test as shown in Fig. \ref{fig:pks}. It is evident from this figure that the {\it noise} bias affects the computation of the cluster position angles negligibly and much less than that of the cluster shapes.
This finding is important since it supports the reliability of observational works where the BCG-Cluster alignment has been detected at high redshift by using a small number of galaxies \citep{west2017}.

\subsection{The Role of Mergers}
\label{subsec:mergers}
%merg_fact        15%                    10%
%Vtang median
%moreProlate      268                   361
%lessProlate      430                   450
%NoAccDt vs Dalpha, all , moreProl
%spearman      0.247158 ,  0.46     0.287125 , 0.545907
%pearson       0.286367 ,  0.54     0.342593 , 0.646133
%KS               0.275862            0.212462
%KS p             0.184429            0.341784

With the aim of assessing the role of major mergers in the evolution of the alignment between the central galaxy and its host cluster we study the individual assembly path of each cluster. Major merger events are defined as accretions of haloes with at least 25\% of the cluster mass. We identify the moments in which a cluster began to accrete another halo as the snapshot just before the accreted halo is last seen as a distinct FOF group. The elapsed time between two successive such moments or, in the case of a last merger the time between it and redshift zero, is defined as \repose.

In this section we concentrate on merger events occurring at $z<1$, where the clusters and the BCGs are more mature, and the average alignment is almost constant.

In order to understand how the BCG-Cluster alignment is affected by major mergers, we consider the change of three quantities between the beginning ($start$) and the end ($end$) of each \repose.

These are:
\begin{itemize}
\vspace{-\topsep}
\item $\Delta\alpha = \alpha_{start}-\alpha_{end}$, where alpha corresponds to the BCG-ClusterDM alignment angle. Positive values of $\Delta\alpha$ indicate an improvement of the alignment during \repose, and vice-versa.
\item $\Delta T = T_{start}-T_{end}$, where $T=(a^2-b^2)/(a^2-c^2)$ is the triaxiality parameter. Values of $T$ near to one (zero) correspond to more prolate (oblate) systems. In turn, positive (negative) values of $\Delta T$ imply that the DM halo is more oblate (prolate) at the end of \repose\ than at the beginning.
\item $\Delta Shift = Shift_{start}-Shift_{end}$, where $Shift$ is the distance between the center of mass and the minimum potential of the cluster. This quantity is often used to characterize the relaxation status of a cluster. Larger values of $\Delta Shift$ indicate that the cluster got more relaxed after \repose, and vice-versa.
\end{itemize}

\begin{figure}
    \includegraphics[width=\columnwidth]{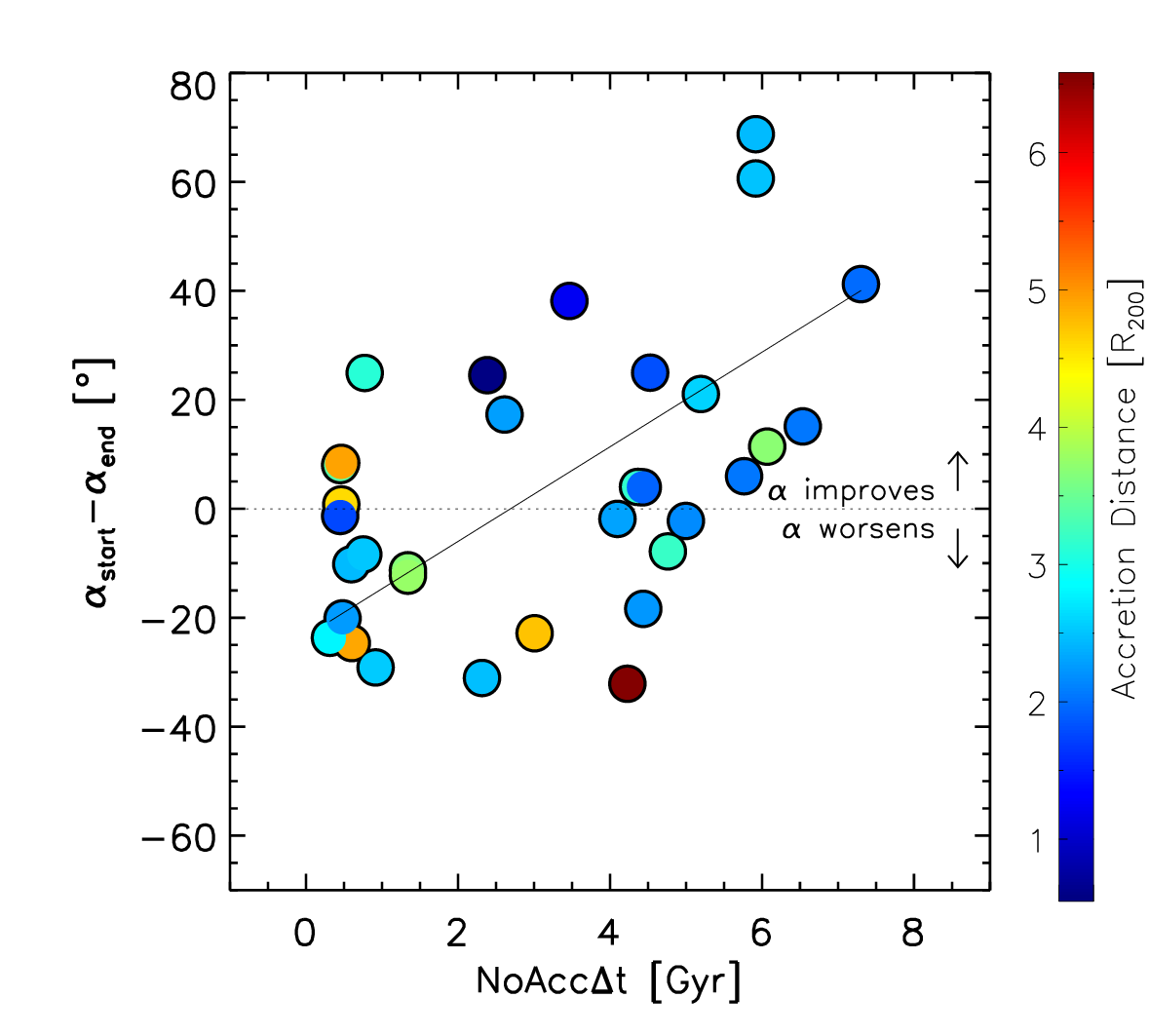} \includegraphics[width=\columnwidth]{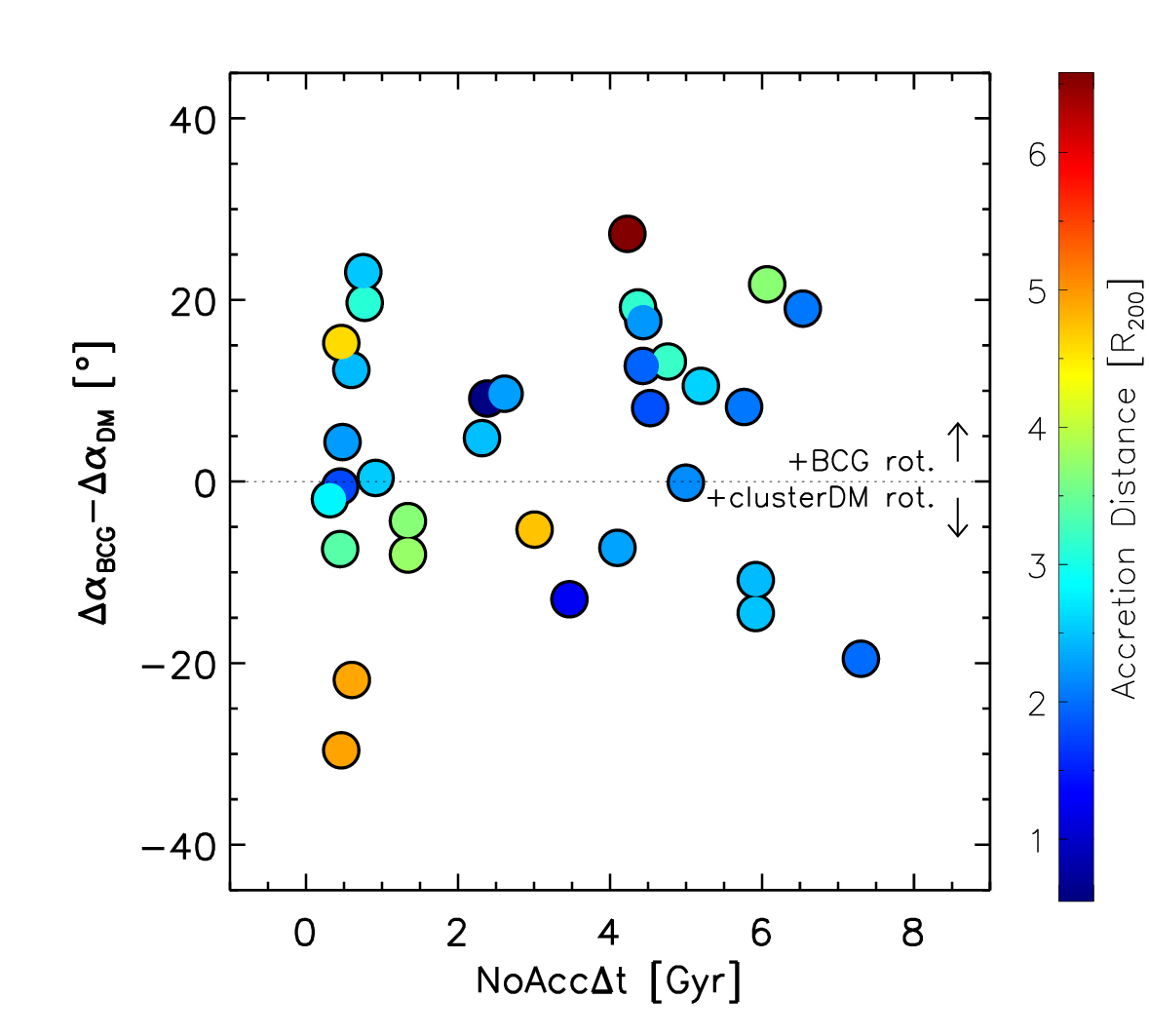}
    \caption{Top: Change in alignment during \repose~ as a function of \repose. Clusters with longer elapsed time between two major mergers have a more evident alignment improvement. The Spearman rank correlation coefficient is $r=0.46$ with a probability of 0.007 of obtaining it from an uncorrelated population. The solid line is  a two-variable linear regression fit to the sample. %corresponding to a 99.9\% confidence level
    %with a significance of $0.007$. The significance is the probability of obtaining $r$ from an uncorrelated population.}
    %Pearson correlation coefficient: 0.50 (33 major mergers)}
    Bottom: The difference between the angular rotation of the BCG and that of the clusterDM principal axes, as a function of \repose. Positive (negative) values mean that the BCG principal axis experienced more (less) rotation than the cluster axis during \repose, as seems to be the case for the longer \repose.
    }
    \label{fig:merg_star_end}
\end{figure}

\begin{figure*}
    \includegraphics[width=\columnwidth]{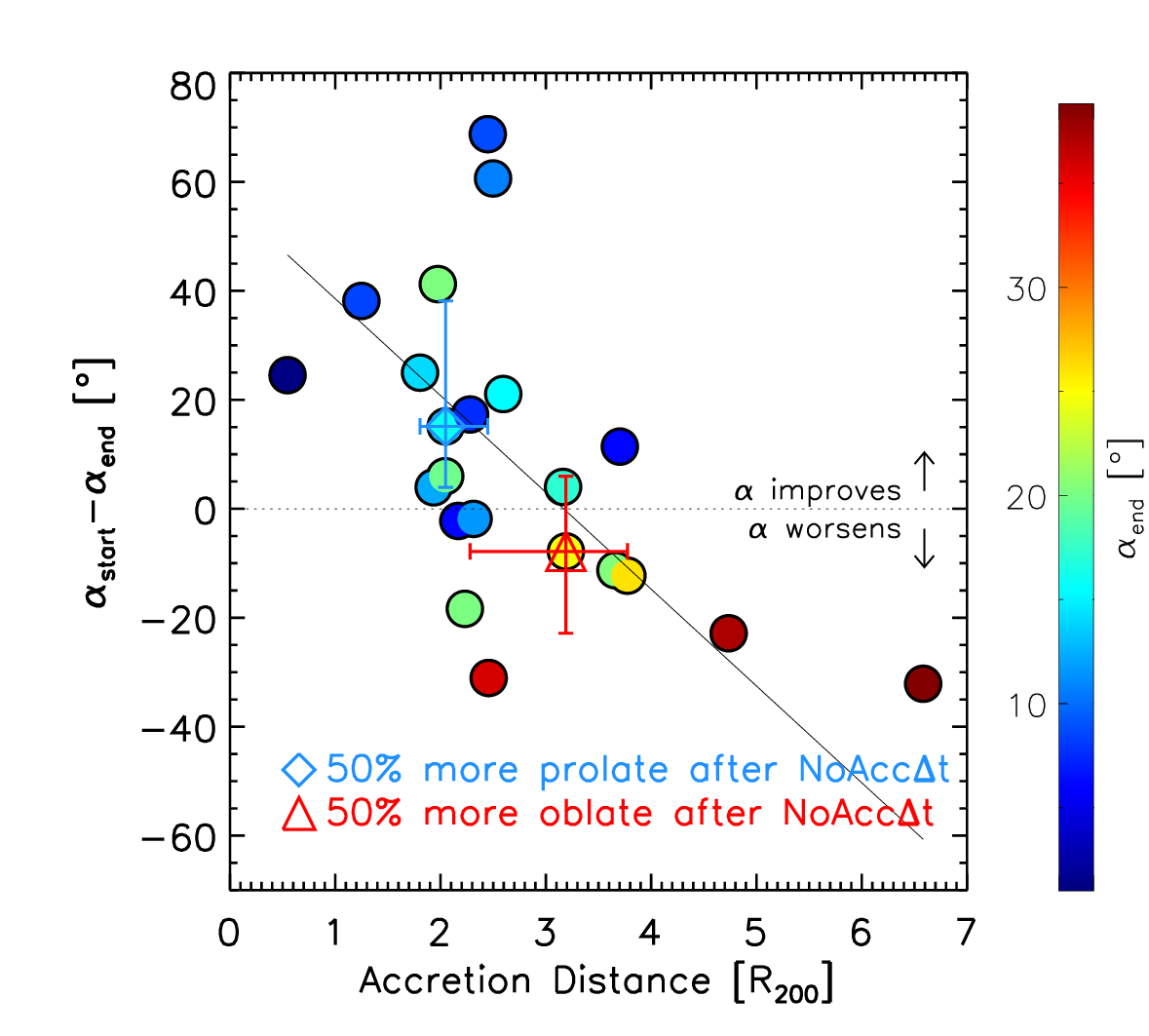}
    \includegraphics[width=\columnwidth]{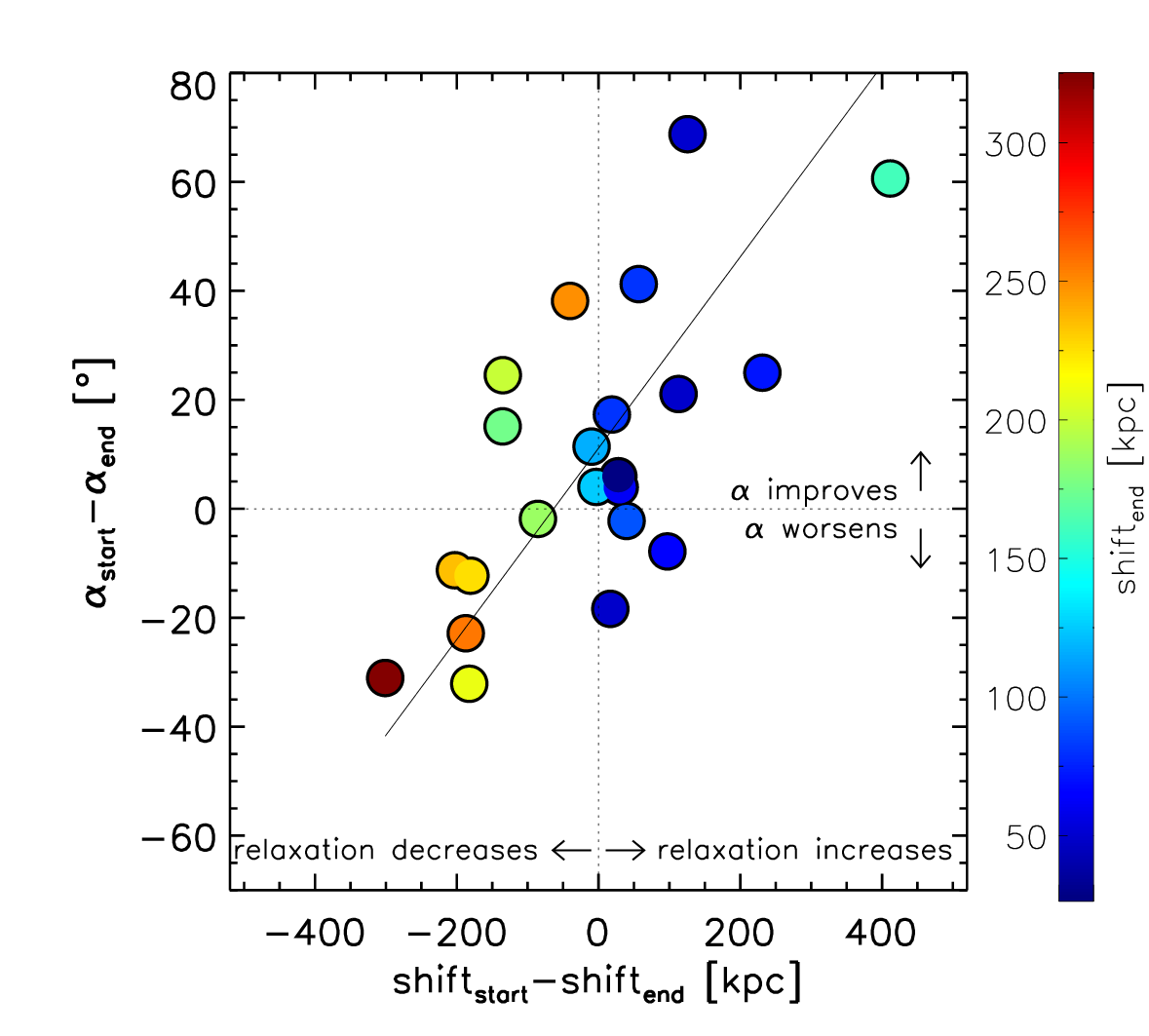}
    \caption{In this plot we consider mergers with \repose\ > 1Gyr. The solid lines are two-variable linear regression fits to the samples. Left: Clusters that improved the alignment with their BCGs have preferentially had major mergers with lower $AccretionDistance$, which means that the accreted halo came from a direction relatively near to the cluster elongation axis. The Spearman rank correlation coefficient is $r=-0.55$ with a probability of $0.009$ of being obtained from an uncorrelated population.
    %Pearson correlation coefficient:-0.54.
    %-0.41 and -0.42 if considering all merger events.
    The empty diamond and triangle stand for the [Median($AccretionDistance$),Median($\Delta\alpha$)] coordinates for the 50\% of clusters  that became more prolate and more oblate, respectively.
    Right: An improvement of the alignment is correlated with an improvement in the cluster relaxation status. The Spearman rank correlation coefficient is $r=0.65$ with a probability of $0.001$ of being obtained from an uncorrelated population.}
    \label{fig:axismergers}
\end{figure*}

For mergers happening at $z<1$ the top panel of Fig. \ref{fig:merg_star_end} shows\footnote{Solid lines in Fig.\ref{fig:merg_star_end} and Fig. \ref{fig:axismergers} are computed with the IDL routine ROBUST\_LINEFIT using the BISECT option.} that the improvement (worsening) of the alignment is more evident when \repose~is longer (shorter). In other words, clusters that spend longer time intervals without important accretion events can strengthen the alignment with their BCGs. This fact could then explain the worsening of the BCG-ClusterDM alignment at the highest redshifts in Fig \ref{fig:alig_evo}, since mergers at those epochs are expected to be more frequent than at late times.
The color code in the figure corresponds to the parameter $Accretion Distance$ which takes into account the geometry of the merger. More precisely, this distance considers the velocity direction and the position that a halo had at the moment of being accreted by the cluster:
\begin{equation}
Accretion Distance=IP+d_p
\end{equation}
where $IP$ is the impact parameter of the accreted halo, hence the perpendicular distance between the direction of its velocity vector and the center of the cluster; and $d_p$ is the perpendicular distance of the accreted halo to the major elongation axis of the cluster.
Mergers entering the cluster near its major axis and with velocities directions nearly parallel to it, {\it on-axis} mergers, will have lower values of $Accretion Distance$. Both $IP$ and $d_p$ are evaluated taking into account the position and relative velocity of the accreting halo just before its last identification as a distinct FOF group.

Coming back to the top panel of Fig. \ref{fig:merg_star_end} it can be noticed a tendency for clusters which improved their alignments to have smaller $Accretion Distance$. In the bottom panel we show  the difference of angular displacements, $\Delta\alpha_{BCG}-\Delta\alpha_{DM}$, of the BCG and clusterDM principal axes during \repose. Positive (negative) values mean that the BCG principal axis experienced more (less) rotation than the cluster axis during \repose. If we take the cases with \repose $> 2 Gyr$, we get that in the 65\% of the cases the BCG principal axis is the one that rotates more toward the new alignment configuration.

In Fig. \ref{fig:axismergers} where we consider only clusters with \repose$>1Gyr$ as we intend to reject systems in which the elapsed time since accretion is comparatively short. The left panel depicts the correlation between $\Delta\alpha$ and  $Accretion Distance$. In this plot we can observe a systematic improvement of the alignment in clusters with smaller $Accretion Distance$, which means clusters accreting material along nearly their principal axes.
The color code in this panel corresponds to the alignment angle at $t_{end}$, the {\it end} of \repose, these angles are always $\lesssim 40^\circ$ with the stronger alignments in clusters with $\Delta\alpha > 0$.  Out of 22 mergers, we have ~85\% (64\%) of the cluster with  $\alpha < 30^\circ$ ($20^\circ$) at $t_{end}$.

We now focus on the shape of clusters at the beginning and at the end of \repose.
We select two subset of clusters according to the median $\Delta T$ defined before which takes into account the change in the triaxiality of the clusters during \repose. The empty diamond in the left panel of Fig. \ref{fig:axismergers} corresponds to the 50\% of clusters with the lowest values of $\Delta T$. For these clusters the median($\Delta T$)= -0.21 which means that at the end of \repose~ they are typically  more prolate than at the beginning.
On the contrary, the 50\% of clusters with the largest values of $\Delta T$, empty triangle in Fig. \ref{fig:axismergers}, has a median($\Delta T$)= 0.13, and hence they became more oblate.
The emerging picture is the following. Clusters that became more prolate after the merger had accretions events typically coming from near the major axis of the cluster and have also improved their alignment. Conversely, those that became more oblate typically had {\it off-axis} acretions and deteriorated their alignments.
%This is in line with Drakos et al. (2019) who state that mergers on radial orbits produce prolate remnants, while mergers on tangential orbits produce oblate remnants.

These are important findings since they suggest that major mergers are not affecting equally the relative orientation of clusters with their central galaxies. Indeed, the frequency and geometry of mergers seem to be related to the final outcome of the BCG-Cluster alignment.
%On the observational side, there has been recent claims suggesting that in the local Universe the BCG-Cluster alignment would be robust against cluster major mergers \citep{wittman2019}.

Finally, the right panel of Fig. \ref{fig:axismergers} is devoted to study the relationship between the alignment and the dynamical state of clusters, as measured by $\Delta Shift$. The color in this panel indicates that clusters with the largest values of $\Delta Shift$ are the more relaxed systems at the {\it end} of \repose.
There is a clear tendency for clusters that became more relaxed to have also the highest improvements in the alignment. This finding is in agreement with Fig \ref{fig:merg_star_end}, since it is naturally expected that clusters with the longest \repose\ have gained a more relaxed status.

\section{Summary and Conclusions}
\label{sec:conclusion}
In this work, we employed cosmological hydro-simulations of rich galaxy clusters to analyse the alignment between their Brightest Cluster Galaxies (BCGs) and the general structure of the clusters, as traced both by the DM distribution and by the distribution of member galaxies.
By reconstructing the main progenitor path of each cluster, we can study the evolution of its alignment angle with the central galaxy. We find that each cluster presents a different alignment angle evolution, which seems to be related to the frequency and geometry of the mergers that the system has experienced.
Depending on their geometry, mergers can promote, destroy or weaken alignments.
If the merger acts in detriment of the alignment, but the cluster is given sufficient time without further important accretions, then the alignment can be restored.
Taking all clusters together, a clear signal of alignment is on average present during the whole studied redshift range.

The main results of this work can be summarized as follow.
\begin{itemize}
\vspace{-\topsep}
\item We find a constant and strong BCG-Cluster alignment signal in the last 10 Gyr ($z \lesssim 2$). The alignment is present whether we define the BCG as star particles inside a fixed aperture of 50kpc or a variable size aperture of 10\%$R_{500}$. The same result holds if we use the DM or the satellite galaxies distributions to obtain the cluster  principal axis.

\item At $z \gtrsim 2$ the BCG-ClusterDM alignment angle increases with redshift, a fact that can be ascribed to the higher frequency of mergers occurring at these epochs. However, the latter behaviour depends on the exact definition of BCG.
In this redshift regimes the definition of BCG becomes increasingly meaningless. Nevertheless, we measure the angle between the proto-cluster and its most massive galaxy.

\item Clusters feature a substantial degree of alignment with the larger-scale structure, as defined by 3$R_{200}$. The same is true for BCGs, albeit to a lesser extent. Taken together, these findings suggest that the alignment is induced from the outside, correlating first a larger-scale with the cluster and then the cluster with its central galaxy.

\item The signal of alignment at $z\lesssim 2$ persists, albeit weakened to some extent, when considering projected matter distributions. The low number galaxies affects less the computation of the cluster principal axis than the computation of its axes ratio.

\item Major mergers may transiently disrupt the alignment. Nevertheless, after some Gyr without further major perturbations, the alignment is developed again. This is accompanied with a more relaxed state for the cluster.

\item Mergers along the cluster principal axis affect the alignment to a lesser extent than off-axis ones.

\item Clusters that after the merger are more prolate than before, improve the alignment more than clusters that became more oblate.
\end{itemize}
Our results suggest a scenario according to which cluster orientations, and consequently on average also BCG orientations, are dictated by the large scale structure. It is indeed conceivable that the relationship between the large scale structure and the orientation of the cluster is produced both by tidal torques and by the preferential direction of accretion and merging onto the cluster. These preferred accretion channels, in turn, affect the orbital parameters of the acquired satellite galaxies, whose interactions and mergers with the BCG will ultimately influence its orientation. Sufficiently relaxed clusters could further orientate the BCG with its gravitational potential through tidal torquing. A detailed analysis of the latter processes will be the subject of future work.

\section*{Acknowledgements}
This project has received funding from the Consejo Nacional de Investigaciones Cient\'ificas y T\'ecnicas de la Rep\'ublica Argentina (CONICET), from the Secretar\'ia de Ciencia y T\'ecnica de la Universidad Nacional de C\'ordoba - Argentina (SeCyT) and from
the European Union's Horizon 2020 Research and Innovation Programme under the Marie Sklodowska-Curie grant agreement No 734374.
Simulations have been carried out CCAD-UNC, which is part of SNCAD-MinCyT (Argentina), and at the computing centre of INAF (Italia). We acknowledge the computing centre of INAF-Osservatorio Astronomico di Trieste, under the coordination of the CHIPP project \citep{bertocco2019,taffoni2020}, for the availability of computing resources and support.
SB acknowledges financial support from PRIN-MIUR 2015W7KAWC, the INFN INDARK grant, the Italy-Germany MIUR-DAAD bilateral grant n. 57396842

%%%%%%%%%%%%%%%%%%%%%%%%%%%%%%%%%%%%%%%%%%%%%%%%%%

%%%%%%%%%%%%%%%%%%%% REFERENCES %%%%%%%%%%%%%%%%%%

% The best way to enter references is to use BibTeX:

\bibliographystyle{mnras}
%\bibliography{bib} % if your bibtex file is called example.bib

% Alternatively you could enter them by hand, like this:
% This method is tedious and prone to error if you have lots of references
%\begin{thebibliography}{99}
%\bibitem[\protect\citeauthoryear{Author}{2012}]{Author2012}
%Author A.~N., 2013, Journal of Improbable Astronomy, 1, 1
%\bibitem[\protect\citeauthoryear{Others}{2013}]{Others2013}
%Others S., 2012, Journal of Interesting Stuff, 17, 198
%\end{thebibliography}

%%%%%%%%%%%%%%%%%%%%%%%%%%%%%%%%%%%%%%%%%%%%%%%%%%

%%%%%%%%%%%%%%%%% APPENDICES %%%%%%%%%%%%%%%%%%%%%

%\appendix

%\section{Some extra material}

%If you want to present additional material which would interrupt the flow of the main paper,
%it can be placed in an Appendix which appears after the list of references.

%%%%%%%%%%%%%%%%%%%%%%%%%%%%%%%%%%%%%%%%%%%%%%%%%%

% Don't change these lines
\bsp	% typesetting comment
\label{lastpage}
\end{document}